\magnification=1200

\centerline{\bf POLARIZATION OF THE CMB PHOTONS}
\centerline{\bf CROSSING THE GALAXY CLUSTERS}
\vskip 2cm
\centerline{\bf{Marina Gibilisco}}\vskip 1mm
\centerline{\it Universit\'a degli Studi di Pavia,}\vskip 1mm
\centerline{\it and INFN, Sezione di Pavia,}\vskip 1mm
\centerline{\it Via Bassi, 6, 27100, Pavia, Italy}\vskip 1mm
\vskip 5mm
\centerline{\it and Dipartimento di Fisica - Universit\'a di Milano}
\centerline{\it Sezione di Astrofisica,}
\centerline{\it Via Celoria, 16, 20133 Milano, Italy.}
\vskip 4mm
\centerline{\it Pacs codes: 98.50 k, 98.70 v, 95.30, 98.80}
\vskip 4mm
\centerline{\bf Abstract:} 
\vskip 4mm
In this paper, I investigate a local effect of polarization
of the Cosmic Microwave Background (CMB) in clusters of galaxies, induced by
the Thomson scattering of an anisotropic radiation.
A local anisotropy of the CMB is produced by some scattering and gravitational
effects, as, for instance, the Sunyaev Zel'dovich effect, the Doppler shift 
due to the cluster motion and the gravitational lensing. 
The resulting anisotropy $\Delta I / I$ depends on the physical properties 
of the clusters, in particular their emissivity in the X band, their size, their
gravitational potential and the peculiar conditions characterizing 
the gas they contain. By solving the Boltzmann radiative transfer equation 
in presence of such anisotropies I calculate the average polarization at the 
centre of some clusters, namely A2218, A576 and A2163, whose properties are 
quite well known. I prove that the gravitational effects due to the 
contraction or to the expansion have some importance, 
particularly for high density structures; moreover, the peculiar motion of 
the cluster, considered as a gravitational lens, influences the 
propagation of the CMB photons by introducing a particular angular dependence 
in the gravitational anisotropy and in the scattering integrals.
Thus, the gravitational and the scattering effects overally produce an 
appreciable local average polarization of the CMB, may be observable through 
a careful polarization measurements towards the centres of the
galaxy clusters.

\vfill\eject

\centerline {\bf 1. INTRODUCTION}
\vskip 4mm
The first detection of the CMB anisotropy, performed by COBE 
(Smoot et al. 1992) four years ago, opened a large debate about the presence 
of a detectable polarization in the background radiation: in fact, one 
expects (Rees 1968) that the Thomson scattering of such an anisotropic 
radiation on the electrons plasma produces a linear polarization. 

The CMB propagation is generally studied by using the Boltzmann transfer 
equation (Peebles 1980); the Chandrasekhar formalism (Chandrasekhar 1960)
is particularly useful to study a polarization problem as the one I want
to handle. After the introduction of a scattering matrix that depends
on the polar angles $(\theta ', \phi ')$ of the incoming beam, the 
evolution of the radiation with the time, the photon energy and with 
the space coordinates is described by using a partial derivative
differential equation, containing, as variables, the four Stokes parameters 
fixing its polarization status at all the times.

One should distinguish between a $global$ and a $local$ effect of 
CMB anisotropy and polarization: the global anisotropy is produced by some
primordial density fluctuations (see, for instance, 
Harari \& Zaldarriaga 1993) and by some time-dependent fluctuations of 
the space-time metric (i.e. by gravitational waves, originated, for 
instance, during the inflationary era (Sachs \& Wolfe 1967). 

I discussed these effects, particularly those due to tensor perturbations, 
in (Gibilisco 1995; Gibilisco 1996); similar approaches to this problem
can be found in (Basko \& Polnarev 1980; Polnarev 1985; Crittenden et al. 1993a
and 1993b).

Here, I want to study the CMB transfer equation from a 
$local$ point of view, i.e. considering the anisotropy effects in clusters
of galaxies: in this case, the clusters properties determine in a
fundamental way the characteristics of the CMB photons which cross
them. In particular, the possible sources of CMB anisotropy in cluster of 
galaxies are the gravitational bound (GB) effect, the collapse (C) or 
expansion (E) effect, the gravitational lensing (GL), the Sunyaev 
Zel'dovich (SZ) effect and finally a Doppler-like shift (D): all these 
effects will be discussed below.

In this work I prove the importance of the gravitational phenomena:
in particular, the motion of the cluster, considered as a lens,
introduces an additional angular dependence in the scattering integral, thus 
conditioning the cancellation of some terms there; the gravitational collapse 
effect dominates for massive clusters, having a high density, while 
the Doppler shift is always negligible if compared to the SZ effect;
note however that it may have some importance in a different framework, namely
in galactic halos: in this case, one expects a total irrelevance of 
the SZ and of the gravitational effects, due to the low values of density and
temperature characterizing the halos; only a significant fraction of MACHOS 
within them could enhance the relevance of the gravitational phenomena,
but more theoretical investigations are necessary to clarify this problem.

Here, through the mathematical solution of the Boltzmann transfer equation,
I calculate the overall CMB polarization induced by the Thomson scattering
of this locally anisotropic radiation and I evaluate the relative importance 
of these effects. In the following, for simplicity, I will assume to have 
a spherical shape for the clusters and an isothermal model 
for the gas they contain (see, for instance, Boynton et al. 1982).
\vskip 4mm
This paper is structured as follows: in sect. 2 I will briefly discuss
the theory of the radiative transfer, by recalling the main equations
governing the photons propagation in presence of Thomson scattering;
I will also recall the formalism of the Stokes parameter describing 
the  polarization status of the CMB. In sect. 3 I will discuss
in some detail the peculiar effects contributing to create a 
local anisotropy for the CMB crossing the clusters;
their dependence on the polar coordinates will be examined and 
the consequences of such a behaviour will be pointed out.
In sect. 4 I will solve the transfer equation by studying the space evolution
of the Stokes parameters on time-constant hypersurfaces; then, I will
show my results for the average polarization degree 
in some clusters, namely A2218, A576 and A2163.
Finally, in sect. 5 I will present my conclusions, emphasizing the 
importance of the gravitational effects in producing a local polarization
of the CMB.
\vskip 4mm

{\bf 2. THE THEORY OF THE RADIATIVE TRANSFER}
\vskip 4mm
A detailed discussion of the radiative transfer theory in presence 
of Thomson scattering was given by Chandrasekhar (Chandrasekhar 1960);
the fundamental equation 
governing the evolution of the photon distribution function 
is the Boltzmann equation (for a detailed explanation see, for instance,
(Peebles 1980)) mathematically expressing the Liouville 
theorem of the conservation of the probability density in the phase
space. 

A suitable way to write the Boltzmann equation adopts as
variables the four Stokes parameters, locally 
describing the polarization status of the radiation (Chandrasekhar 1960);
these parameters are used as components of a 4-vector 
$ n_{\alpha}=n_{\alpha}(\theta,\phi,\nu)$, which is
a function of polar angles $(\theta, \phi)$ and of the photon frequency $\nu$:
$$
n_{\alpha}~\equiv~(I_{l},~I_{r},~U,~V).\eqno(2.1)
$$
In eq. (2.1) $I_{l},~I_{r}$ are the left and right intensities, 
defined as in the Chandrasekhar geometrical formalism (Chandrasekhar 1960)
based on the polarization ellipse (see fig. 1);
with this definition, the total intensity of the radiation is
$I~=~I_{l}+I_{r}$, while $Q~=~I_{l}-I_{r}$ and $U$ are the parameters
indicating the presence of a linear polarization; finally, $V$ represents the
circular polarization, but its evolution can be neglected because
it totally decouples from the others parameters (see Appendix). As a result, 
the Thomson scattering of an anisotropic radiation
produces $linear$ polarization only, as proved by Rees in (Rees 1968).

The final expression for the radiative transfer
is a partial differential equation that reads:
$$
\Big(~{\partial \vec n\over \partial \eta}~+~\gamma^{\alpha}~
{\partial\vec n \over
{\partial x^{\alpha}} }~\Big)~+~{\partial\nu \over {\partial\eta}}~
{\partial\vec n\over{\partial\nu}}~=~{\sigma_{T}~N_{e}~R(\eta)~\over 4\pi}~
\times
$$
$$
\times \Big [~-4\pi\vec n~+~\int^{1}_{-1}~\int^{2\pi}_{0}~\vec n(\mu ',\phi ')
~P(\mu,\phi,\mu ',
\phi ')~d\mu ' ~d\phi '~\Big].\eqno(2.2)
$$
In eq. (2.2), $\eta$ is the comoving time, defined as $\int [dt/R(t)]$,
and $R(t)$ is the scale factor of the Universe;
$\gamma^{\alpha}$ are the components of an unit vector in the propagation 
direction of the photons, $\sigma_{T}=6.65~\times~10^{-25}$
$cm^{2}$ is the Thomson scattering cross-section, $\mu~=~\cos \theta$ and 
$P(\mu,\phi,\mu ',\phi ')$ is the Chandrasekhar scattering matrix, whose 
explicit form as a function of the polar angles is given in (Chandrasekhar 1960)
and partially shown in the Appendix; finally, $N_{e}$ is the comoving number 
density of free electrons, depending on the ionization history of the Universe.

An analytical way to solve the transfer equation in presence of 
cosmological gravitational waves has been discussed in (Gibilisco, 1995)
on the basis
of the theory of the Volterra integral equations: here it is sufficient to 
recall that eq. (2.2) contains a collisionless part, given by the
time component of the geodesic equation, and a collisional term,
represented by the integral in its right-hand side. That really
gives all the informations about the scattering processes,
responsible for the production of a linearly polarized, outgoing radiation.

\vskip 4mm
{\bf 3. LOCAL EFFECTS PRODUCING THE CMB ANISOTROPY AND POLARIZATION
IN CLUSTERS OF GALAXIES}
\vskip 4mm
Two kinds of local effects in galaxy clusters may produce an anisotropy 
and, consequently, a polarization in the CMB crossing these
structures: the first class consists of gravitational effects,
due to the distortion of the Hubble flow in the region of
a massive clusterl; the latter one is due both to the inverse
Compton scattering of the photons on the plasma electrons
and to a Doppler-like shift, connected with the motion of the structure.
In the following I will recall in some detail all these effects,
pointing out the local CMB anisotropy they induce in order to determine
the level of linear polarization generated by the Thomson scattering.

\vskip 4mm

\item{a)} {\bf Gravitational effects:}
\vskip 4mm
\item{1)} {\it Moving Gravitational lenses:} 

\vskip 4mm

A massive object always influences the radiation that propagates
in the surrounding region,
gravitationally distorting its path; in particular, a source 
observed behind a massive structure appears slightly different
in frequency and flux if compared to a similar one located in 
a different position: that represents a typical lensing phenomenon.

A similar, but less known, effect is produced when the object representing
the gravitational lens is {\it in motion} across the line of sight
(Birkinshaw 1989); in this case, the lens produces a) an {\it anisotropy} in 
the CMB crossing the lens and b) a redshift difference between
the multiple images of a background object seen through it.

These effects have been discussed in (Birkinshaw 1989; Birkinshaw \& Gull
1983); here we are interested in the production of an anisotropy in the CMB, 
a phenomenon that depends on the transverse velocity of the lens
and presents a peculiar angular dependence.

Looking at fig. 2a, we consider the CMB as seen through a convex
lens (for instance a galaxy cluster), moving with a velocity $\vec v$
which forms an angle $\alpha$ with the line of sight: $\delta$ is the deflection
angle of the radiation and $\theta$ is the angle between the deflected 
direction of propagation and the line of sight (see fig. 2b) 
(Birkinshaw \& Gull 1983).
In the lens frame, the frequency of the radiation is unchanged: 
on the contrary, for a stationary observer in the frame of the original
source, the observed frequency of the deflected light depends 
on $\vec v$ and $\delta$ as follows (Birkinshaw \& Gull 1983):
$$
{\Delta \nu\over \nu}~=~
\gamma~\beta~\delta~\sin\alpha~\cos\phi;
\eqno(3.1)
$$
here $\beta$ and $\gamma$ respectively are $v/c$ and the Lorentz factor, 
while $\phi$ is the angle between the lens 
velocity vector and the emerging ray projected onto the sky (see fig. 2b).
The deflected photons have different energies and different 
brightness in comparison with the undeflected ones: as a result, the
corresponding change in the brightness temperature of the radiation
field is (Birkinshaw \& Gull 1983):
$$
{\Delta T\over T}(\theta, \phi,\alpha)~=~\Bigg(~{8\gamma\beta G M_{0}\sin\alpha
\over c^{2}L}~\Bigg)~\cos\phi\cdot {\theta_{0}\over \theta}~\Big[
1-\big(1-{\theta ^{2}\over \theta_{0}^{2}}\Big)^{3/2}~\Big];\eqno(3.2)
$$
eq. (3.2) holds for $\theta\leq\theta_{0}$; $M_{0}$ is the lens mass
(that, for a galaxy cluster, is about $10^{15}~M_{\odot}$),
$L$ is the average cluster radius ($L\sim 3~Mpc$) and $\theta_{0}$
is given by the ratio $L/d$ where $d$ is the cluster distance.

From eq. (3.2) one can remark that the CMB anisotropy 
and the induced polarization could be relevant for cluster of galaxies 
having large transverse peculiar velocities: in fact, 
in a rich cluster of galaxies, the deflection angle due to the lensing
is relatively large ($\sim 1~ arcmin$), while the transverse
velocity may be larger than 1000 $km~s^{-1}$ (Dressler 1987).

\vskip 4mm

\item{2)} {\it Effects due to the gravitational bound:}

\vskip 4mm

The decoupling of a massive, gravitationally bounded structure
from the local Hubble flow produces a redshift 
in the radiation that crosses it, when compared to the one that
passes far away; in fact, the metric of the massive object changes
in a different way as regards the one of the surrounding Universe. 
As a result, a monochromatic source observed behind 
a massive structure will be observed at a shifted frequency compared to
a similar source placed in a different position. 

This redshift effect in a variable gravitational field was theoretically studied
by(Rees \& Sciama 1968; Dyer 1976; Nottale 1984)
in the framework of a ``Swiss cheese'' model: 
this model starts from a zero-pressure Robertson-Walker 
Universe where one removes a comoving sphere of dust and places
a clump of the same mass at the centre of the ``hole'' thus obtained.
In this way, the optical effects of the massive object are influenced 
by the surrounding void in the embedding Universe: that is 
due to the localized changes of the space-time 
metric in the region where the inhomogeneity is present and where the
distortion of the Hubble flow is significant.

In this model, a cluster is represented by a high density Friedmann
solution of the Einstein equations, separated by an empty Schwarzschild
zone from the surrounding, lower density Friedmann Universe (Nottale 1984);
the radiation crossing both the empty Schwarzschild zone and the 
massive central structure suffers two effects (Dyer 1976):
a) a redshift effect, implying that the 
parameters $1+z~=~\nu / \nu_{0}$ and $1+\xi~=~R(T_{0})/ R(T)$
($R$ is the scale factor) do not coincide behind the cluster;
b) a time-delay effect, due to the fact that the radiation 
crossing the inhomogeneous structure follows a longer path.
For the CMB photons the temperature anisotropy resulting from 
these effects reads (Nottale 1984):
$$
{\Delta T\over T}~=~-4~q_{0}~{\overline{\rho_{c}}\over \rho_{0}}~
\Big ({H_{0}r\over c}\Big)^{3}~\Big(~1-{H_{c}\over H_{0}}-\ln 
\Big({\overline{\rho_{c}}\over \rho_{0}}\Big)^{1/3}~\Big),\eqno(3.3)
$$
where $q_{0}$ is the deceleration parameter,
$L$ is the cluster size, $\overline{\rho_{c}}$ and $\rho_{0}$
are the average density of the cluster and the one of the background Universe
and finally $H_{c}$ and $H_{0}$ are the Hubble constants for the cluster
and for the background Universe at the present time.
The ratio $\overline{\rho_{c}} /\rho_{0}$ is about equal to 10 in the case
of an expanding cluster, while it is about 1000 for a collapsing structure 
(Nottale 1984).
\vskip 4mm

\item{3)} {\it Effects due to the gravitational collapse or expansion:}

\vskip 4mm

A phenomenon very similar to the one above discussed is produced 
when the change of the metric is due to the
collapse or to the expansion of the massive object (Nottale 1984): a rich 
cluster may both contract ($H_{c}/H_{0} <0$) or expand ($H_{c}/H_{0} >0$); 
however, the observations seem to support the possibility of a collapse 
(Capelato et al. 1982; de Vaucouleurs 1982).

The resulting temperature anisotropy of the CMB photons is (Nottale 1984):

\item {a)} for a contracting cluster ($ H_{c}/H_{0} <0$):
$$
{\Delta T\over T}~=~-5~\Bigg[~q_{0}~{\overline{\rho_{c}}\over \rho_{0}}~
\Big ({H_{0}r\over c}\Big)^{2}~\Bigg]^{3/2};\eqno(3.4)
$$
\item {b)} for a slightly expanding cluster ($ 0< H_{c}/H_{0} \leq 1$):
$$
{\Delta T\over T}~=~-4~q_{0}~{\overline{\rho_{c}}\over \rho_{0}}~
\Big ({H_{0}r\over c}\Big)^{3}~\Big(q_{0}~{\overline{\rho_{c}}\over 
\rho_{0}}-\ln \Big({\overline{\rho_{c}}\over \rho_{0}}\Big)^{1/3}~\Big);
\eqno(3.5)
$$
\item{c)} for a rapidly expanding cluster ($H_{c}/H_{0} >1$):
$$
{\Delta T\over T}~=~4~q_{0}~{\overline{\rho_{c}}\over \rho_{0}}~
\Big ({H_{0}r\over c}\Big)^{3}~{H_{c}\over H_{0}};\eqno(3.6)
$$
These gravitational effects seem to dominate 
the scattering ones; that is surely true for very massive clusters
($M>10^{19}~M_{\odot}$), as proved in (Rees \& Sciama 1968);
in the following I will prove
their relevance also for lower masses ($M\sim 10^{14}-10^{15}~M_{\odot}$).

\vskip 4mm

\item{b)} {\bf The Sunyaev-Zel'dovich and the Doppler effects:}

\vskip 4mm

The Sunyaev-Zel'dovich effect (SZ) (Sunyaev \& Zel'dovich 1972)
consists in a characteristic distortion of the CBM photons spectrum due 
to the inverse Compton scattering of the radiation
on the hot electron gas ($T\sim 10^{8}~K$) 
present in the clusters (Lea et al. 1973).
That produces a shift in the CMB spectrum towards
smaller wavelengths near the blackbody peak at 
$\lambda\sim 1~mm$ and a consequent decrement of the radiation intensity
in that region; correspondingly, one observes an increment of the intensity 
at submillimetric wavelengths. 

This effect has been studied at centimeter wavelengths in many
clusters, as Abell 2319 (White \& Silk 1980) Abell 576 (Boynton et al. 1982),
(White \& Silk 1980), Abell 2218 (Boynton et al. 1982; 
Klein et al. 1991; Jones et al. 1993), Abell 665 and Cl0016+16 
(Birkinshaw et al. 1984); a significant evidence of the
SZ distortion has probably been obtained for Abell 2218 (Boynton et al. 1982).
In fact, the properties of the core gas in A576 and in some other 
clusters desagree with the theoretical expectations for the inverse
Compton scattering process and therefore the apparent CMB decrement 
could be due to some spurious effects (White \& Silk 1980). 
Anyway, the presumable average 
magnitude of the temperature decrement is in the range 
$3\times 10^{-5}\div 10^{-3}~K^{\circ}$ (White \& Silk 1980); for Abell 2218
Klein et al. found $\Delta T=-0.21\pm 0.04~mK^{\circ}$ 
at $\lambda=1.2~cm$ (Klein et al. 1991), while Jones et al. (Jones et al. 1993)
found $\Delta T\sim 0.5~mK^{\circ}$.

From a theoretical point of view, the evaluation of the magnitude of
the SZ effect depends on the cluster model one assumes, in particular
on the $X$ ray emissivity, on the cluster size and mass, on the form of the 
gravitational potential and on the physical conditions characterizing 
the core gas. 
Generally, one supposes the hot gas is in an isothermal and quasi-hydrostatic 
equilibrium and the cluster is a stationary and spherically symmetric 
structure (White \& Silk 1980).

The $X$-ray emission mainly comes from the bremsstrahlung processes 
and it is given by (White \& Silk 1980):
$$
\Lambda (T)~=~3.2\times 10^{-24}\Bigg[~\ln~\Big(1+{0.13T\over {E_{1}+E_{2}}}
\Big)~+~2.51\Big({{E_{1}+E_{2}}\over T}~+~0.5\Big)^{-1/2}~\Bigg]\times
$$
$$
\times ~T^{1/2}~\Big[~exp(-E_{1}/T)~-~exp(-E_{2}/T)~\Big]~ergs~cm^{3}~s^{-1},
\eqno(3.7)
$$
where the passband is $(E_{1},E_{2})$ and $T$ is the gas temperature,
all expressed in $KeV$. The emissivity profile is given by (White \& Silk 1980):
$$
{L(r)}~=~n_{e}^{2}(r)~\Lambda(T)~=~{n_{e}^{2}(r)\over 2}~{l_{0}\over r_{c}}~
\Big(1~+~{r^{2}\over r_{c}^{2}}\Big)^{-3/2},\eqno(3.8)
$$
where $l_{0}$ and $r_{c}$ are the central surface brightness
and the $X$-ray core radius, known from the observations.

Finally, in an isothermal model for the gas (Boynton et al. 1982; Klein et al. 
1991) the density profile is given by the following formula:
$$
n_{e}~=~n_{e,0}~\Big(1~+~{r^{2}\over r_{c}^{2}}\Big)^{-3n/2};\eqno(3.9)
$$
where $n_{e,0}$ is the central electron density, equal to $(3.4\pm 1.5)
\times 10^{-3}~cm^{-3}$, and $n$ is the power law index, equal to 
$0.5\pm 0.1$ (Klein et al. 1991).

For a gas whose properties are given by eqs. (3.6)-(3.9), the 
microwave decrement at the centre of the cluster in the Rayleigh-Jeans limit 
is (White \& Silk 1980):
$$
{\Delta T_{M}\over T_{M}}=-{2k\sigma_{T}\over m_{e}c^{2}}~(2r_{c}l_{0})^{1/2}~
\cdot \int^{\infty}_{0}~T(x)(\Lambda [T(x)])^{-1/2}~(1+x^{2})^{-3/4}~dx,
\eqno(3.10)
$$
where $\sigma_{T}=6.65\times 10^{-25}~cm^{2}$ is the Thomson scattering 
cross section and the microwaves temperature has been called $T_{M}$ in order 
to distinguish it from the gas temperature $T$; finally, $x=h\nu/kT$. 

In an isothermal model with $T=12.5~KeV$ the integration of 
eq. (3.10) gives (White \& Silk 1980):
$$
{\Delta T_{M}\over T_{M}}=-7.4~{k\sigma_{T}\over m_{e}c^{2}}~(r_{c}l_{0})^{1/2}~
T~[\Lambda (T)]^{-1/2};\eqno(3.11)
$$
eq. (3.11) is the final expression I will assume for the SZ temperature 
decrement.
\vskip 4mm
An additional decrement is due to the peculiar motion of the cluster 
(Sunyaev \& Zel'dovich 1972 and 1980):
that is just a Doppler-like effect and it causes a change of the radiation 
intensity and of the CMB temperature depending on the radial component 
$v_{r}$ of the peculiar velocity of the cluster (Sunyaev \& Zel'dovich 1980):
$$
{\Delta T_{M}\over T_{M}}~=~-{v_{r}\tau\over c},\eqno(3.12)
$$
$$
{\Delta I\over I}~=~-{x~exp(x)\over (exp(x)-1)}~{v_{r}\tau\over c};
\eqno(3.13)
$$
here $\tau$ is the optical depth, equal to $\tau=\int^{L}_{-L}~n_{e}(r)
\sigma_{T}~dr$.

While the thermal SZ effect gives the same perturbation both
for the temperature and the intensity, in this case the expressions
for $\Delta T/ T$ and $\Delta I/I$ are different: in particular, 
the perturbed intensity only (eq. (3.13)) depends on the frequency.
This effect should be dominant in galactic halos, where the gravitational and 
thermal contributions probably are negligible.
\vskip 4mm

{\bf 4. THE CMB POLARIZATION IN A2218, 
A576, A2163}.
\vskip 4mm
Now I can apply the formalism developed in sects. 2 and 3
to study the CMB polarization in some observed clusters.

The clusters considered in this analysis were studied in many 
observations and their properties are quite well determined; moreover,
they are possible candidates for which the observation of the 
Sunyaev-Zel'dovich decrement has been claimed.

In tabs. 1a, 1b and 1c I resume the physical properties of these clusters
useful for my calculation of the CMB polarization.
\vskip 4mm
{\bf 4.1 THE CLUSTERS A2218, A576, A2163.}
\vskip 4mm

{\bf Cluster A2218}

Following the original Abell catalogue, 
A2218 is classified as a richness class 4, distance class 6 cluster.
Subsequently, its redshift has been fixed to 0.174 (Le Borgne et al. 1992)
and its class fixed as
Bautz-Morgan II (Leir \& Van den Bergh 1978); it contains a large cD galaxy
and many weak radio sources (Andernach et al. 1988); 
a review of the present characteristics of 
A2218 can be found in (Boynton et al. 1982; Klein et al. 1991).

\vskip 4mm
{\bf Cluster A576}
\vskip 4mm
A576 is a moderately rich cluster, classified as Abell richness class 1,
distance class 2 and Bautz-Morgan class III 
(Leir \& Van den Bergh 1978); it presents a central
condensation region but none dominant galaxy; the cluster contains a
high proportion of SO galaxies (Melnick \& Sargent 1977), 
it is surrounded by an extensive $X$-ray emitting halo (Forman et al. 1978) 
but it is a quite weak $X$-ray source.
A review of its properties can be found in (White \& Silk 1980).

\vskip 4mm
{\bf Cluster A2163}
\vskip 4mm
A2163 is a rich cluster of Rood and Sastry class I (Struble \& Rood 1987) (
corresponding to $N_{Abell}=119$ galaxies) having a redshift $z=0.201$.
The spectroscopic $X$-ray observations, performed with the GINGA satellite 
(Arnaud et al. 1992) 
showed an exceptionally high temperature and $X$-ray luminosity 
($kT=13.9^{+1.1}_{-1.0}~KeV$ and $L_{X}=6.0\times 10^{45}~
erg/s$ in the band $2-10~KeV$, while the usual values for other clusters
are respectively lower than $9~KeV$ and $2\times 10^{45}~erg/sec$).
Due to the fact that the SZ effect is proportional to the product of 
the electron density and of the temperature, A2163 surely represents
a very promising candidate for the observation of the SZ microwave decrement.
A review of the present experimental knowledge of A2163 can 
be found in (Elbaz et al. 1995).

\vskip 4mm
{\bf 4.2 THE CALCULATION OF THE POLARIZATION}
\vskip 4mm
\noindent Before calculating the CMB polarization induced by the 
previously discussed anisotropies we should firstly fix a suitable reference 
frame to work. If we look at figs. 2a and 2b, a possible choice consists
in fixing the origin of the frame in coincidence with the observer on the earth;
the $z$ axis is oriented towards the centre of the 
cluster, along the line of sight. For simplicity, I assume an approximate 
spherical symmetry for the clusters, a hypothesis in most cases 
well confirmed by the observations.

I introduce the usual polar coordinate system
$(r,\theta,\phi)$ and I define $\theta_{0}=L/d$, where $L$ is 
the cluster radius and $d$ is its distance from the observer.
Then, I write the transfer equation (2.2) in this particular reference frame.
Before the scattering Thomson the anisotropic radiation 
is unpolarized: the perturbed intensity reads 
$$
[\Delta I]_{tot}~=~[\Delta I(\theta,\phi,\alpha)]_{Lens}~+~
[\Delta I(r)]_{Grav. b. + 
coll./exp.}~+~[\Delta I(T)]_{SZ}~+~[\Delta I(\nu,v_{r})]_{Doppl},\eqno(4.2.1)
$$
i.e. it is given by the sum of the various contributions previously discussed.
The incoming radiation in the transfer equation is represented by a vector
$\vec n_{in}$ containing the perturbed intensity:
$$
\vec n_{in}~\equiv~(\Delta I/2,~\Delta I/2,~0).\eqno(4.2.2)
$$
(as usual, I neglect the Stokes parameter $V$ because no circular 
polarization is produced).
After the Thomson scattering on the freee eleectron plasma,
the anisotropic radiation acquires a linear polarization (Rees 1968) and the
vector expressing its status is:
$$
\vec n_{scatt}~\equiv~(\Delta I_{l},~\Delta I_{r},~\Delta U).\eqno(4.2.3)
$$
Then, the transfer equation in polar coordinates reads:
$$
3~{\partial \vec n_{scatt}\over \partial r}~+~{1\over r}\cdot
{\partial \vec n_{scatt}\over \partial \theta}~[2\tan\theta~-~\cot\theta]
~+~{1\over r}\cdot{\partial \vec n_{scatt}\over \partial \phi}~
[\tan\phi~-~\cot\phi]~=~
$$
$$
-\sigma_{T}~n_{e}(r)~R(\overline{t})~\vec n_{scatt}~+~{\sigma_{T}\over 4\pi}~
R(\overline{t})~n_{e}(r)~\int^{2\pi}_{0}\int_{\mu_{0}}^{1}
~d\mu '~d\phi '~P(\mu,\mu ',\phi, \phi ')~
\vec n_{in};\eqno(4.2.4)
$$
($\mu_{0}=\cos\theta_{0}$; the primed angular variables refer to the 
incoming radiation).

Eq. (4.2.4) holds on fixed-time hypersurfaces: 
the reference time $\overline t$ corresponds to the observed redshift of 
the cluster. The variable $\theta$ in the integration spans over the range
$(0,\theta_{0})$.
\vskip 4mm

{\bf 4.3 THE RESULTS}
\vskip 4mm
The CMB polarization is obtained by solving eq. 
(4.2.4): due to the axial simmetry of the problem, this expression 
simplifies, because the term containing the $\phi$ derivative 
vanishes.

Substituting the vectors (4.2.2) and (4.2.3) into eq. (4.2.4),
I obtain a system of three partial differential equations
in the perturbed parameters $\Delta I_{l}$, $\Delta I_{r}$, $\Delta U$,
whose solution represents the final polarization status of the CMB 
in the cluster.
The system reads as follows:
$$
3~{\partial \Delta I_{l}\over \partial r}~+~{1\over r}~
{\partial \Delta I_{l}\over \partial \theta}~(2\tan\theta~-~\cot\theta)~=~
\sigma_{T}~n_{e}(r)~R(\overline{t})~\Big[~{I_{0}\over 8\pi}~
R_{1}(\theta,\phi,r,\nu,\alpha, T)~-~\Delta I_{l}~\Big],\eqno(4.3.1a)
$$
$$
3~{\partial \Delta I_{r}\over \partial r}~+~{1\over r}~
{\partial \Delta I_{r}\over \partial \theta}~(2\tan\theta~-~\cot\theta)~=~
\sigma_{T}~n_{e}(r)~R(\overline{t})~\Big[~{I_{0}\over 8\pi}~
R_{2}(\theta,\phi,r,\nu,\alpha, T)~-~\Delta I_{r}~\Big],\eqno(4.3.1b)
$$
$$
3~{\partial \Delta U\over \partial r}~+~{1\over r}~
{\partial \Delta U\over \partial \theta}~(2\tan\theta~-~\cot\theta)~=~
\sigma_{T}~n_{e}(r)~R(\overline{t})~\Big[~{I_{0}\over 8\pi}~R_{3}
(\theta,\phi,r,\nu,\alpha, T)~-~\Delta U~\Big].\eqno(4.3.1c)
$$
In eqs. (4.3.1a), (4.3.1b) and (4.3.1c) I called $R_{1},~R_{2}$ and $R_{3}$
the scattering integral that appears in eq. (4.2.4), while $R(t)$
is the scale factor of the Universe; I separate
the different contributions to the anisotropy in the following way:
$$
\Big( {\Delta I\over I}\Big)_{tot}~=~F_{1}(\theta ',\phi ',\alpha)~+~
F_{2}(r)~+~F_{3}(\nu,T),\eqno(4.3.2)
$$
where $F_{1}$ is given by eq. (3.2) and represents the lensing contribution
(the primed angular variables must be used because we are referring to the
incoming radiation) $F_{2}$ is given by eq. (3.3) plus one of 
the eqs. (3.4), (3.5) or (3.6), depending on the cluster evolution
(gravitational bound + collapse or expansion term)
and finally $F_{3}$ is given by the sum of eq. (3.11) (thermal SZ effect)
and eq. (3.13) (Doppler effect); then the scattering integrals read 
($j=1,2,3$):
$$
R_{j}(\theta,\phi,r,\nu,\alpha, T)~=~\int^{2\pi}_{0}~\int^{1}_{\cos\theta_{0}}~
d\mu '~d\phi '~(a_{j1}+a_{j2})~\Big[~F_{1}(\theta ', \phi ',\alpha)~+~
F_{2}(r)~+~F_{3}(\nu,T)~\Big].\eqno(4.3.3)
$$
In eq. (4.3.3), $a_{j1}$ and $a_{j2}$ are the matrix elements
of the Chandrasekhar scattering matrix, whose form can be found in 
Appendix.

A possible way to perform the calculation consists in searching 
a recursive solution of eqs. (4.3.1a), (4.3.1b) and (4.3.1c): I expand the
perturbations in Legendre polynomials as follows:
$$
\Delta I_{l}~=~{I_{0}\over r}~\sum_{l}~a_{l}~P_{l}(\cos\theta),
\eqno(4.3.4a)
$$
$$
\Delta I_{r}~=~{I_{0}\over r}~\sum_{l}~b_{l}~P_{l}(\cos\theta),
\eqno(4.3.4b)
$$
$$
\Delta U~=~{I_{0}\over r}~\sum_{l}~c_{l}~P_{l}(\cos\theta),
\eqno(4.3.4c)
$$
where $I_{0}$ is the unperturbed intensity, given by:
$$
I_{0}~=~{1\over {exp(h\overline \nu /k_{B}T)-1}};\eqno(4.3.5)
$$
here $h\overline \nu\sim 10^{-13}~GeV$ and $k_{B}T\sim 2.35\times 10^{-13}~
GeV$.
Then, it is possible to calculate the coefficients of the expansion
simply by solving the differential equations (4.3.1a), (4.3.1b)
and (4.3.1c); the recursive calculation can be stopped at $l=2$, because
the higher order terms in the polynomials are negligible.
Finally, the polarization degree $P$ is
$$
P(r,\theta,\phi)~=~{|~\Delta I_{l}-\Delta I_{r}+\Delta U~|\over 
~(\Delta I_{l}+\Delta I_{r})}.\eqno(4.3.6)
$$
In tabs. 2a, 2b and 2c I show a comparison of the different contributions
$F_{1},~F_{2}$ and $F_{3}$ to the local CMB anisotropies for the considered 
clusters and in the cases of a contracting, slightly expanding or totally 
expanding cluster; the data listed in these tables have been calculated 
at some reference values for the polar coordinates and for an angle 
$\alpha=\pi/2$ in order to maximize the lensing contribution.
In tab. 3 I list the values of the coefficients which appear
in the expansions (4.3.4a), (4.3.4b) and (4.3.4c) of the perturbed Stokes 
parameters for the three clusters: 
I show the values obtained for fixed $(r,\theta,\phi)$
in order to give an idea of the size of these coefficients.
Finally, in figs. 3, 4 and 5 I plotted the average polarization degree in the 
clusters A2218, A576 and A2163 as resulting from this calculation.
\vskip 4mm
{\bf 5. DISCUSSION AND CONCLUSIONS}
\vskip 4mm
Looking at eqs. (4.3.1a), (4.3.1b) and (4.3.1c), we remark that 
the different evolution of the perturbed 
Stokes parameters originates in the scattering integrals, as an effect of
the dissimilarity of the functions $R_{1},~R_{2}$ and $R_{3}$.

The angular dependence of the matrix elements $a_{j1},~a_{j2}$
(see eqs. (A.1), (A.2) and (A.3) of the Appendix) and of the
gravitational lensing term $F_{1}$ plays a fundamental 
r\^ole in determining the Stokes parameters evolution and, therefore,
in producing a CMB polarization; for instance, the $R_{3}$ scattering integral 
vanishes when integrated on $\phi '$: as a result, all the coefficients
$c_{l}$ of the expansion (4.3.4c) are equal to zero and the overall
contribution of $\Delta U$ to the polarization cancels.
A linear polarization results from the scattering contributions $R_{1}$ 
and $R_{2}$: the difference existing between these functions causes an 
unequal evolution of the perturbed left and right intensities and 
therefore an unvanishing term $|\Delta I_{l}-\Delta I_{r}|$ in eq. (4.3.6).

In tabs. 2a, 2b and 2c, I considered both the possibilities of a collapse or 
an expansion for the clusters: we can remark that the most important 
contribution to the CMB anisotropy and polarization is due to
the gravitational collapse of the clusters ($F_{2}$ term); the high densities 
attained in this case strongly enhance the importance of the gravitational 
effects but, anyway, also for an expanding cluster $F_{2}$ dominates.
Moreover, the SZ effect ($F_{3}$ contribution)
dominates over the lensing one ($F_{1}$ contribution), while
the Doppler shift is totally negligible at the considered frequencies;
note however that the lensing contribution slightly
influences the angular behaviour of the scattering integral, by introducing 
an additional $\cos \phi '$ dependence.

In tab. 3 I listed the coefficients of the expansion
in Legendre polynomials, obtained by solving the differential equations 
(4.3.1a), (4.3.1b) and (4.3.1c): the recursive calculation can be stopped 
at the order $l=2$ because the coefficients vanish very fast. 

These results confirm that the gravitational effects are important 
in determining the characteristics of the CMB photons crossing the galaxy
clusters, particularly in the case of very massive and dense structures.
In figs. 3, 4 and 5  I showed the behaviour of the average polarization degree
in galaxy clusters as a function of the polar angle $\theta$: due to the
smallness of $\theta_{max}=\theta_{0}=L/d$, this dependence is very weak,
indicating that an observation performed towards the clusters centres
should be able to measure the CMB polarization induced by these effects.
Possible fluctuations in the signal might be attributable to the presence
of some central radiosources which introduce a serious source of noise.

The average polarization degree for A2218 and A2163 is comparable, while
for A576 it is smaller by a factor 10: that is probably due to the 
$z$ dependence of the gravitational bound effect.

No relevant differences are found between the polarization in expanding 
or in collapsing clusters: that is due to the fact that a change in the 
function $F_{2}$ affects in the same way the values of the left and right
perturbed intensities, thus leaving unchanged the quantity
$|\Delta I_{l}-\Delta I_{r}|$.
\vskip 4mm
Summarizing, in this paper I studied the CMB polarization in 
clusters of galaxies induced by local anisotropies: by 
expanding the Stokes parameters in Legendre polynomials
and by solving the Boltzmann transfer equation through a recursive 
method, I proved that the gravitational effects have some importance: 
they indeed influence the properties of the CMB photons
crossing the galaxy clusters.
The cancellation of some terms in the Thomson scattering integral
makes $|\Delta I_{l}-\Delta I_{r}|$ different from zero,
while the resulting perturbation in the Stokes parameter $U$ vanishes.

In such a way, the Thomson scattering of the locally anisotropic CMB 
radiation (the anisotropy being due both to the scattering and to the
gravitational effects) produces an appreciable polarization in clusters
having a mass $M\geq 10^{14}\div 10^{15}~M_{\odot}$
An accurate observation towards the clusters centres might, in principle, 
reveal such a phenomenon, provided that one is able to distinguish the 
contribution due to the possible presence of strong radiosources.

\vfill\eject
\centerline{{\bf APPENDIX}}
\vskip 4mm
The Chandrasekhar scattering matrix (Chandrasekhar 1960) is a $4\times 4$ 
matrix having the following structure:
$$
\left (
\matrix { a_{11}  &  a_{12} & a_{13}  & 0 \cr
a_{21}  &  a_{22} & a_{23}  & 0 \cr
a_{31}  &  a_{32} & a_{33}  & 0 \cr
0       &   0     &   0     &  a_{44} \cr}
\right)
$$
The presence of only one matrix element different from zero in the position 
(4,4) assures the Stokes parameter $V$ is totally
decoupled from the other ones: as a result, the Thomson scattering 
of the anisotropic radiation does not produce a circular polarization.

When we apply the scattering matrix to the vector (4.2.2), representing
the incoming radiation, we need to know the following linear combinations
of the matrix elements $a_{ij}$:

for eq. (4.3.1a):
$$
a_{11}~+~a_{12}~=~{3\over 4}~\Bigg[~3\mu^{2}\mu^{'2}~-~\mu^{2}~-~2\mu^{'2}
~+~2~+~4~\sqrt{1-\mu^{2}}~\sqrt{1-\mu^{'2}}~\mu\mu^{'}~\cos(\phi' -\phi)~+
$$
$$
+~\mu^{2}\mu^{'2}\cos 2(\phi'-\phi)-\mu^{2}\cos 2(\phi'-\phi)~\Bigg];\eqno(A.1)
$$
for eq. (4.3.1b):
$$
a_{21}~+~a_{22}~=~{3\over 4}~\Bigg[ ~1~+~\mu^{'2}~+~\cos 2(\phi'-\phi)~
(1-\mu^{'2})~\Bigg];\eqno(A.2)
$$
for eq. (4.3.1c):
$$
a_{31}~+~a_{32}~=~{3\over 2}~\Bigg[~ -2\mu^{'}
\sqrt{1-\mu^{2}}~\sqrt{1-\mu^{'2}}~\sin (\phi'-\phi)~+~\sin 2(\phi'-\phi)~
(\mu-\mu\mu^{'2})~\Bigg].\eqno(A.3)
$$
The $\phi '$ dependence of eqs. (A.1), (A.2) and (A.3) produces the cancellation
of some terms in the scattering integrals; the anisotropy due to the 
lensing effect (see eq. (3.2)) also contains a factor $\cos \phi'$,
thus, combining these dependences, I obtain from eq. (4.3.3) 
$R_{3}=0$ and $R_{1}\not= R_{2}$. This difference between $R_{1}$
and $R_{2}$ is finally responsible for the creation of a local
polarization in clusters.

\vskip 4mm
\centerline{{\bf ACKNOWLEDGEMENTS}}
\vskip 4mm
I am particularly grateful to the University of Pavia not only for 
the financial support given to my research but also for its
pleasant hospitality, for the courtesy of the people that 
helped me in many way and for the technical support 
given to my work; in particular,
I am grateful to Bruno Bertotti for some discussions about the problems
connected with the polarization of the Cosmic Microwave Background.

A grateful thank goes also to all the people of the Astrophysical Section
of the University of Milano; in particular, I would like to address
an especial thank to Silvio Bonometto, who firstly gave me the idea to
investigate about this problem and then the way to pursue this work.
Finally, I want to address the most affectionate thank to my 
friends, Emma and Francesco.

\vskip 1cm
\centerline {{\bf REFERENCES}}
\vskip 4mm

\item{} Andernach, H., Tie, H., Sievers, A., Reuter, H.P., Junkes, N.,
Wielebinski, R., {\it Astron. and Astroph. Suppl.}, 1988, {\bf 73}, 265.

\item{} Arnaud, M., Hughes, J.P., Forman, W., Jones, C., Lachieze-Rey,
M., Yamashita, K., Hatsukade, I., {\it Astroph. Journ.}, 1992, {\bf 390}, 345.

\item{} þBasko, M.M., Polnarev, A.G., {\it Sov. Astr.}, 1980, {\bf 24}, 268.

\item{} Birkinshaw, M., {\it Measurement of the Sunyaev Zel'dovich effect},

\item{} Birkinshaw, M., Gull, S.F., {\it Nature}, 1983, {\bf 302}, 315.

\item{} Birkinshaw, M., Gull, S.F., Hardebeck, H.E., {\it Nature}, 1984,
{\bf 309}, 34.

\item{} Birkinshaw, M., 1989, {\it Gravitational Lenses}
Eds. Moran, Hewitt and Lo, Springer Verlag, Berlin, p.59.

\item{} Boynton, P.E., Radford, S.J.E., Schommer, R.A., Murray, S.S.,
{\it ýAstroph. Journ.}, 1982, {\bf 257}, 473.

\item{} Capelato, H.V. et al., {\it Astroph. Journ.}, 1982, {\bf 252}, 433.

\item{} Chandrasekhar, S., 1960, {\it Radiation Transfer}, Dover, New York.

\item{ } Crittenden, R., Bond, J.R., Davis, R.L.,Efstathiou, G.,
Steinhardt, P.J., {\it Phys. Rev. Lett.}, 1993a, {\bf 71}, 324.

\item { } Crittenden, R., Bond, J.R, Davis, R.L., Efstathiou, G.,
Steinhardt, P.J., {\it Phys. Rev. Lett.}, 1993b, {\bf 69}, 1856.

\item{} de Vaucouleurs, G., {\it Astroph. Journ.}, 1982, {\bf 253}, 520.

\item{} Dressler, A. et al., {\it Astroph. Journ.}, 1987, {\bf 313}, L37.

\item{} Dyer, C.C., {\it MNRAS}, 1976, {\bf 175}, 429.

\item{} Elbaz, D., Arnaud, M., B\"orhinger, H.,
{\it Astron. and Astroph.}, 1995, {\bf 293}, 337.

\item{} Forman, W., Jones, C., Murray, S.S., Giacconi, R.,
{\it Astroph. Journ.}, 1978, {\bf 225}, L1.

\item {} Gibilisco, M., {\it Int. Journ. of Mod. Phys.}, 1995, {\bf 10A}, 
3605.

\item {} Gibilisco, M., {\it Astroph. and Space Science}, 1996, {\bf 235}, 75.

\item{} Harari, D.D., Zaldarriaga, M., {\it Physics Lett.}, 
1993, {\bf B319} ,96.

\item{} Jones, M. et al., {\it Nature}, 1993, {\bf 365}, 320.

\item{} Klein, U., Rephaeli, Y., Schlickeiser, R., Wielebinski, R.,
{\it Astronomy and Astrophys.}, 1991, {\bf 244}, 43.

\item{} Lea, S.M., Silk, J.I., Kellogg, E., Murray, S.S, 
{\it Astroph. Journ.}, 1973, {\bf 184}, L105.

\item{} Le Borgne, J.F., Pello, R., Sanahuja, B., {\it Astron. and Astroph.
Suppl.}, 1992, {\bf 95}, 87.

\item{} Leir, A.A., Van den Bergh, S., {\it Astroph. Journ. Suppl.}, 1978,
{\bf 34}, 381.

\item{} Melnick, J., Sargent, W.L.W., {\it Astroph. Journ.}, 1977,
{\bf 215}, 401.

\item{} Nottale, L., {\it MNRAS}, 1984, {\bf 206}, 713.

\item{} P.J.E. Peebles, 1980, {\it Large Scale Structures in the Universe},
Princeton University Press.

\item{ } Polnarev, A.G., {\it Sov. Astron.}, 1985, {\bf 29}, 607.

\item{} Rees, M. J., {\it Astroph. Journ.}, 1968, {\bf 153}, L1.

\item{} Rees, M.J., Sciama, D.W., {\it Nature}, 1968, {\bf 217}, 511.

\item{} Sachs R.K., Wolfe, A.M., {\it Astroph. Journ.}, 1967, {\bf 147}, 73.

\item{} Serlemitsos, P., Smith, B., Boldt, E., Holt, S., Swank, J.,
{\it Astroph. Journ.}, 1977, {\bf 211}, L63.

\item{} Smoot, G.F. et al., {\it Astroph. Journ.}, 1992, {\bf 396}, L1.

\item{} Struble, M.F., Rood, H.J., {\it Astroph. Journ. Suppl.}, 1987,
{\bf 63}, 555.

\item {} Sunyaev, R.A., Zel'dovich, Y.B., {\it Comm. Astr. Sp. Phys.}, 
1972, {\bf 4}, 173.

\item{} Sunyaev, R.A., Zel'dovich, Y.B., {\it MNRAS}, 1980, {\bf 190}, 413.

\item{} White, S.D.M., Silk, J.I., {\it Astroph. Journ.}, 1980, {\bf 241}, 864.

\vfill\eject
{\bf Tab. 1a: Properties of the cluster Abell 2218 (Boynton et al. 1982;
Klein et al. 1991).

\vskip 0.5cm
{\offinterlineskip
\tabskip=0pt
\halign{ \strut
	 \vrule#&
\quad    \bf# 
              \hfil \quad    &
	 \vrule#&
\quad	 \hfil #  &
	 \vrule#
	 \cr
\noalign{\hrule}
&~~~~~~~~~                     &&~                                  &\cr
& CLUSTER PROPERTIES~~           && DATA                            &\cr
&                                &&                                 &\cr
\noalign{\hrule}
&                                &&                                 &\cr
&    $z$                         &&  $0.174$                        &\cr  
&                                &&                                 &\cr
&    $d$                         &&  $1060~\pm~ 810~Mpc$            &\cr  
&                                &&                                 &\cr
&    $L$                         &&  $\sim~3~Mpc$                   &\cr  
&                                &&                                 &\cr
&    $\theta_{0}$                &&  $2.83\times 10^{-3}$           &\cr
&                                &&                                 &\cr 
&    $r_{c}$                     &&  $0.22~\pm~0.06~Mpc$            &\cr  
&                                &&                                 &\cr
&    $l_{0}$                     &&  $1.60\times 10^{-4}~erg~cm^{-2}~sec^{-1}$
                                                                    &\cr
&                                &&                                 &\cr
&    $n_{0}$                     &&  $(4.2~\pm~1.8)\times 10^{-3}~cm^{-3}$
                                                                    &\cr
&                                &&                                 &\cr
&    $x$                         &&  $0.438$ ($\nu=25~GHz$)         &\cr
&                                &&                                 &\cr
&    $M_{T}$                     &&  $(7.8~\pm~1.4)\times 10^{14}~M_{\odot}$
                                                                    &\cr
&                                &&                                 &\cr
&    $\beta$                     &&   $0.02$                        &\cr
&                                &&                                 &\cr
&    $\sigma$                    &&   $1400~\pm~200~Km/s$           &\cr
&                                &&                                 &\cr
&    $n$                         &&   $0.5$                         &\cr  
&                                &&                                 &\cr
\noalign{\hrule}
 }}
\vskip 4mm
Note: $d$ = cluster distance; $L$ = cluster radius; $\theta_{0}=L/d$;
$r_{c}$ = X-ray core radius; $l_{0}$ = central surface brightness; $n_{0}$ =
central electron density; $x=h\nu/kT_{r}$; $M_{T}$ = total cluster mass;
$\beta=v_{clu}/c$; $\sigma$ = line of sight velocity dispersion of the galaxies.
$n$ = power law index of the electron density profile.

\vfill\eject
{\bf Tab. 1b: Properties of the cluster Abell A576 (White \& Silk 1980)}

\vskip 0.5cm
{\offinterlineskip
\tabskip=0pt
\halign{ \strut
	 \vrule#&
\quad    \bf# 
              \hfil \quad    &
	 \vrule#&
\quad	 \hfil #  &
	 \vrule#
	 \cr
\noalign{\hrule}
&~~~~~~~~~                     &&~                                  &\cr
& CLUSTER PROPERTIES~~           && DATA                            &\cr
&                                &&                                 &\cr
\noalign{\hrule}
&                                &&                                 &\cr
&    $z$                         &&  $0.039$                        &\cr  
&                                &&                                 &\cr
&    $d$                         &&  $236~Mpc$                      &\cr  
&                                &&                                 &\cr
&    $L$                         &&  $\sim~3~Mpc$                   &\cr  
&                                &&                                 &\cr
&    $\theta_{0}$                &&  $0.013$                        &\cr
&                                &&                                 &\cr 
&    $r_{c}$                     &&  $0.24~Mpc$                     &\cr  
&                                &&                                 &\cr
&    $l_{0}$                     &&  $3.8\times 10^{-5}~erg~cm^{-2}~sec^{-1}$
                                                                    &\cr
&                                &&                                 &\crÿ
&    $n_{0}$                     &&  $4.6~\times 10^{-3}~cm^{-3}$
                                                                    &\cr
&                                &&                                 &\cr
&    $x$                         &&  $0.18$ ($\nu~=~11~ GHz$)       &\cr
&                                &&                                 &\cr
&    $M_{T}$                     &&  $(2.9~\pm~0.5)\times 10^{15}~M_{\odot}$
                                                                    &\cr
&                                &&                                 &\cr
&    $\beta$                     &&   $0.02$                        &\cr
&                                &&                                 &\cr
&    $\sigma$                    &&   $1124~Km/s$                   &\cr
&                                &&                                 &\cr
&    $n$                         &&   $0.64$                        &\cr  
&                                &&                                 &\cr
\noalign{\hrule}
 }}
\vfill\eject
{\bf Tab. 1c: Properties of the cluster Abell 2163 (Struble \& Rood 1987;
Arnaud et al. 1992; Elbaz et al. 1995)}

\vskip 0.5cm
{\offinterlineskip
\tabskip=0pt
\halign{ \strut
	 \vrule#&
\quad    \bf# 
              \hfil \quad    &
	 \vrule#&
\quad	 \hfil #  &
	 \vrule#
	 \cr
\noalign{\hrule}
&~~~~~~~~~                     &&~                                  &\cr
& CLUSTER PROPERTIES~~           && DATA                            &\cr
&                                &&                                 &\cr
\noalign{\hrule}
&                                &&                                 &\cr
&    $z$                         &&  $0.201$                        &\cr  
&                                &&                                 &\cr
&    $d$                         &&  $1267~Mpc$                     &\cr  
&                                &&                                 &\cr
&    $L$                         &&  $\sim~3~Mpc$                   &\cr  
&                                &&                                 &\cr
&    $\theta_{0}$                &&  $2.37\times 10^{-3}$           &\cr
&                                &&                                 &\cr 
&    $r_{c}$                     &&  $0.305~\pm~0.019~Mpc$          &\cr  
&                                &&                                 &\cr
&    $l_{0}$                     &&  $2.16\times 10^{-3}~erg~cm^{-2}~sec^{-1}$
                                                                    &\cr
&                                &&                                 &\cr
&    $n_{0}$                     &&  $6.8\times 10^{-3}~cm^{-3}$    &\cr
&                                &&                                 &\cr
&    $x$                         &&     $0.18$ ($\nu~=~11~GHz$)     &\cr
&                                &&                                 &\cr
&    $M_{T}$                     &&  $(4.6~\pm~0.4)\times 10^{15}~M_{\odot}$
                                                                    &\cr
&                                &&                                 &\cr
&    $\beta$                     &&   $0.02$                        &\cr
&                                &&                                 &\cr
&    $\sigma$                    &&   $Unknown$                     &\cr
&                                &&                                 &\cr
&    $n$                         &&  $0.62^{+0.05}_{-0.02}$         &\cr
&                                &&                                 &\cr
\noalign{\hrule}
 }}
\vfill\eject

{\bf Tab. 2a: Different contributions to the CMB anisotropy and polarization in
A2218.}

\vskip 0.5cm
{\offinterlineskip
\tabskip=0pt
\halign{ \strut
	 \vrule#&
\quad    \bf# 
              \hfil \quad    &
	 \vrule#&
\quad	 \hfil #  &
	 \vrule#&
\quad	 \hfil #  &
	 \vrule#&
\quad	 \hfil # &
	 \vrule#
	 \cr
\noalign{\hrule}
&~~~~~~~~~  ~~~~~~~~~~~ &&              &&~~~~~~~~~~         &&             &\cr
& $F_{1}$               && $F_{2}$(GB)  && $F_{2}$(CO/EXP)   && $F_{3}$     &\cr
&~~~~~~                 &&~~~~~~~~~~~~  &&~~~~~~~~~~~~~~~~~  &&             &\cr
\noalign {\hrule}
&                       &&                  &&               &&             &\cr
&                       && $28.690$         && $-2484.92$ (CO)&&            &\cr
&                       &&                  &&               &&             &\cr
& $1.63\times 10^{-6}$  && $0.116$          && $-0.4720$ (SE)  
                                            $-3.923\times 10^{-4}$          &\cr
&                       &&                  &&               &&             &\cr
&                       && $0.116$          && $0.142$ (EXP) &&             &\cr
&                       &&                  &&               &&             &\cr
  \noalign{\hrule}
 }}
\vskip 7mm

$F_{1}$: Lensing contribution;

$F_{2}$: Gravitational bound contribution (GB); Gravitational collapse 
(CO) or expansion (Slight expansion (SE), Expansion (EXP)) contributions;

$F_{3}$: Sunyaev-Zel'dovich contribution (SZ); the Doppler shift contribution 
is negligible.

The listed values correspond to the following choice of the cluster coordinates
and properties: $r=1063 Mpc$, $\theta=2.83\times 10^{-3}~rad$, $\phi=0.785~rad$,
$\alpha=1.570~rad$.
\vfill\eject
{\bf Tab. 2b: Different contributions to the CMB anisotropy and polarization in
A576.}

\vskip 0.5cm
{\offinterlineskip
\tabskip=0pt
\halign{ \strut
	 \vrule#&
\quad    \bf#  \quad
              \hfil     &
	 \vrule#&
         \quad
	 \hfil #  \quad &
	 \vrule#&
         \quad	 
         \hfil #  \quad &
	 \vrule#&
	 \quad
	 \hfil #  \quad &
	 \vrule#
	 \cr
\noalign{\hrule}
&~~~~~~~~~  ~~~~~~~~~~~ &&              &&~~~~~~~~~~         &&             &\cr
& $F_{1}$               && $F_{2}$(GB)  && $F_{2}$(CO/EXP)   && $F_{3}$     &\cr
&~~~~~~                 &&~~~~~~~~~~~~  &&~~~~~~~~~~~~~~~~~  &&             &\cr
\noalign {\hrule}
&                       &&                  &&               &&             &\cr
&                       && $0.299$          && $-125.62$ (CO)&&             &\cr
&                       &&                  &&               &&             &\cr
& $5.89\times 10^{-6}$  && $1.047\times 10^{-3}$
                                            && $-5.361\times 10^{-3}$ (SE)  
                                            && $-3.183\times 10^{-5}$       &\cr
&                       &&                  &&               &&             &\cr
&                       && $1.047\times 10^{-3}$
                                            &&  $1.342\times 10^{-3}$ (EXP)  
                                                             &&             &\cr
&                       &&                  &&               &&             &\cr
  \noalign{\hrule}
 }}
\vskip 7mm

$F_{1}$: Lensing contribution;

$F_{2}$: Gravitational bound contribution (GB); Gravitational collapse 
(CO) or expansion (Slight expansion (SE), Expansion (EXP)) contributions;

$F_{3}$: Sunyaev-Zel'dovich contribution (SZ); the Doppler shift contribution 
is negligible.

The listed values correspond to the following choice of the cluster coordinates
and properties: $r=239 Mpc$, $\theta=0.013~rad$, $\phi=0.785~rad$,
$\alpha=1.570~rad$.
\vfill\eject
{\bf Tab. 2c: Different contributions to the CMB anisotropy and polarization in
A2163.}

\vskip 0.5cm
{\offinterlineskip
\tabskip=0pt
\halign{ \strut
	 \vrule#&
\quad    \bf# 
              \hfil \quad    &
	 \vrule#&
\quad	 \hfil #  &
	 \vrule#&
\quad	 \hfil #  &
	 \vrule#&
\quad	 \hfil # &
	 \vrule#
	 \cr
\noalign{\hrule}
&~~~~~~~~~  ~~~~~~~~~~~ &&              &&~~~~~~~~~~         &&             &\cr
& $F_{1}$               && $F_{2}$(GB)  && $F_{2}$(CO/EXP)   && $F_{3}$     &\cr
&~~~~~~                 &&~~~~~~~~~~~~  &&~~~~~~~~~~~~~~~~~  &&             &\cr
\noalign {\hrule}
&                       &&                  &&               &&             &\cr
&                       && $49.773$         && $-3546.93$ (CO) &&           &\cr
&                       &&                  &&               &&             &\cr
& $9.56\times 10^{-6}$  && $0.206$          && $-0.804$  (SE)  
                                            $-5.411\times 10^{-5}$          &\cr
&                       &&                  &&               &&             &\cr
&                       && $0.206$          && $0.250$ (EXP) &&             &\cr
&                       &&                  &&               &&             &\cr
  \noalign{\hrule}
 }}
\vskip 7mm

$F_{1}$: Lensing contribution;

$F_{2}$: Gravitational bound contribution (GB); Gravitational collapse
(CO) or expansion (Slight expansion (SE), Expansion (EXP)) contributions;

$F_{3}$: Sunyaev-Zel'dovich contribution (SZ); the Doppler shift contribution 
is negligible.

The listed values correspond to the following choice of the cluster coordinates
and properties: $r=1270 Mpc$, $\theta=2.37\times 10^{-3}~rad$, $\phi=0.785~rad$,
$\alpha=1.570~rad$.

\vfill\eject

{\bf Tab. 3: The coefficients of the expansion in Legendre Polynomial of the 
perturbed Stokes parameters: $a_{l}$ refers to $\Delta I_{l}$, $b_{l}$ to 
$\Delta I_{r}$ while $c_{l}$, referring to $\Delta U$, are zero for any $l$;
the values have been calculated for a fixed r, respectively r=1063 Mpc for
A2218, r=239 Mpc for A576 and r=1270 Mpc for A2163; I chose also 
$\theta=\theta_{0}$ and $\phi=\pi/4$.}

\vskip 0.5cm
{\offinterlineskip
\tabskip=0pt
\halign{ \strut
	 \vrule#&
\quad    \bf# 
              \hfil \quad    &
	 \vrule#&
\quad	 \hfil #  &
	 \vrule#&
\quad	 \hfil #  &
	 \vrule#&
\quad	 \hfil #  &
         \vrule#&
\quad	 \hfil #  &
	 \vrule#
	 \cr
\noalign{\hrule}
&~~~~~~~~~           &&~~~~~~~~~~~ ~           &&~~~~~~~                    &\cr
& CLUSTER            &&   $a_{l}$              &&  $b_{l}$                  &\cr
&                    &&                        &&                           &\cr
\noalign {\hrule}
&~~~~~~              &&~~~~~~~~~~~~            &&~~~~~~~~~~~~~~~~~          &\cr
&        && $a_{0}=-1.206\times 10^{+25}$ && $b_{0}=-7.654\times 10^{+24}$  &\cr
&        &&                               &&                                &\cr
& A2218  && $a_{1}=-7.638\times 10^{+17}$ && $b_{1}=-7.636\times 10^{+17}$  &\cr
&        &&                               &&                                &\cr
&        && $a_{2}\sim 0.0$               && $b_{2}\sim 0.0$                &\cr
&        &&                               &&                                &\cr
\noalign {\hrule}
&        &&                               &&                                &\cr
&        && $a_{0}=-2.801\times 10^{+24}$ && $b_{0}=-2.750\times 10^{+24}$  &\cr
&        &&                               &&                                &\cr
& A576   && $a_{1}=-1.717\times 10^{+17}$ && $b_{1}=-1.738\times 10^{+17}$  &\cr
&        &&                               &&                                &\cr
&        && $a_{2}\sim 0.0$               && $b_{2}\sim 0.0$                &\cr
&        &&                               &&                                &\cr
\noalign {\hrule}
&        &&                               &&                                &\cr
&        && $a_{0}=-1.440\times 10^{+25}$ && $b_{0}=-6.895\times 10^{+24}$  &\cr
&        &&                               &&                                &\cr
& A2163  && $a_{1}=9.126\times 10^{+17}$  && $b_{1}=4.563\times 10^{+17}$   &\cr
&        &&                               &&                                &\cr
&        && $a_{2}\sim 0.0$               && $b_{2}\sim 0.0$                &\cr
&        &&                               &&                                &\cr
\noalign {\hrule}
}}
\vfill\eject

{\bf FIGURE CAPTIONS}

\vskip 4mm
\item {Fig. 1:} The polarization ellipse (see also (Chandrasekhar 1960), p.26);
the principal axes of the ellipse form the angles $\chi$, $\chi +\pi/2$
with the direction $\vec l$; $\psi$ is the angle formed
by the generical vibration direction with $\vec l$. The Stokes 
parameters are defined as a function of $I_{l},~I_{r},~\chi$ and $\beta$
(the tangent of $\beta$ is the ratio of the axes of the
ellipse traced by the end point of the electric vector) as in (Chandrasekhar 
1960).
\vskip 3mm
\item {Fig. 2a:} A moving lens produces a change in the brightness of an 
isotropic radiation field, proportional to the transverse velocity of the 
lens: the deflection causes a slight decrease in the photon energy if
$\theta > 0$ and a slight increase if $\theta < 0$; in the Rayleigh-Jeans
part of the CMB spectrum this effect appears as a brightness
increase or decrease. The figure is taken from (Birkinshaw \& Gull 1983).

\vskip 3mm
\item {Fig. 2b:} The configuration of the problem in the frame of the 
observer; the lens has a velocity $\vec v$ in the $(x,z)$ plane, while the
angles $(\theta,\phi)$ describe the direction of the deflected photon
and respectively correspond to the angle between $\vec k$ and the $z$ axis
and to the angle between the projected $\vec k$ and the projected $\vec v$.
The figure is taken from (Birkinshaw \& Gull 1983).

\vskip 3mm
\item {Fig. 3} The average polarization degree as a function of the 
polar angle $\theta$ in the cluster A2218.

\vskip 3mm
\item {Fig. 4} The average polarization degree as a function of the 
polar angle $\theta$ in the cluster A576.

\vskip 3mm
\item {Fig. 5} The average polarization degree as a function of the 
polar angle $\theta$ in the cluster A2163.
\end